\documentclass[seceq,preprint]{ptptex}

\def\nn{\nonumber}
\def\B{{\rm B}}
\def\IIS{{\rm IIS}}
\def\VEV#1{\left\langle{#1}\right\rangle}
\def\slash#1{{\ooalign{\hfil/\hfil\crcr$#1$}}}
\newcommand{\kslash}{\slash{k}}
\newcommand{\pslash}{\slash{p}}

\preprintnumber[3cm]{
OU-HET584\\ YITP-07-55\\ September, 2007}

\title{
The BV Master Equation for the Wilson Action \\
in general Yang-Mills Gauge Theory%
}

\author{
Takeshi \textsc{Higashi}$^{1,}$\footnote{%
E-mail address:\  {\tt higashi@het.phys.sci.osaka-u.ac.jp}}, %
Etsuko \textsc{Itou}$^{2,}$\footnote{%
E-mail address:\  {\tt itou@yukawa.kyoto-u.ac.jp}} and %
Taichiro \textsc{Kugo}$^{2,}$\footnote{%
E-mail address:\  {\tt kugo@yukawa.kyoto-u.ac.jp}}%
}

\inst{
$^{1}$Department of Physics, 
Graduate School of Science, Osaka University,\\ 
Toyonaka, Osaka 560-0043, Japan \\
$^{2}$
Yukawa Institute for Theoretical Physics, Kyoto University, \\
Kyoto 606-8502, Japan
}

\abst{
The Wilson effective action for general Yang-Mills gauge theory is 
shown to satisfy the usual form of Batalin-Vilkovisky (BV) master equation, 
despite that a momentum cutoff apparently breaks the gauge invariance. 
In the case 
of Abelian gauge theory, in particular, it actually deduces the 
Ward-Takahashi identity for Wilson action recently derived by Sonoda. }

\begin{document}

\maketitle

\section{Introduction}

Exact renormalization group (ERG)\cite{Polchinski:1983gv} 
provides us with a powerful tool to 
reveal the dynamics of various field theories. 
For the important cases of gauge theories, however, 
the very notion of the momentum cutoff to define 
the Wilson action is apparently incompatible with the 
gauge invariance, and there have been many proposals and trials for 
circumventing the difficulty.\cite{Warr:1986,Becchi:1996an,%
Reuter:1993kw,Ellwanger:1994iz,Arnone:2005vd}

In this respect it was truly remarkable that 
Sonoda\cite{Sonoda:2007dj} has recently 
found that a simple form of Ward-Takahashi (WT) identity holds for 
the Wilson action 
in  QED. It implies that the momentum cutoff is compatible with gauge 
invariance. Moreover the Sonoda's WT identity for the Wilson action was 
rederived by a simpler path-integral method by 
Igarashi, Itoh and Sonoda (IIS) in Ref.~\citen{Igarashi:2007fw}. 
Those authors also showed that Sonoda's equation can be lifted 
into the form of quantum Batalin-Vilkovisky (BV) master equation\cite{Igarashi:2001mf}, 
remarkably, exactly the same 
form of equation as the continuum theory without cutoff.

Their work was, however, restricted to QED. In this short note we show that 
the Wilson effective action for a general non-Abelian gauge theory satisfies 
the quantum 
master equation. We follow the method developed by IIS which is 
really powerful and makes it remarkably easy to derive the master equation. 
Again the master equation is written in exactly the same 
form as the usual one for the continuum theory. This derivation is done 
in Sects. 2 and 3.

In the special case of QED, we can make it explicit how the Wilson action 
depends on the BV antifields since the Faddeev-Popov (FP) ghost fields 
are free there. 
Then we shall elucidate the relation of our Wilson master action 
with that derived by IIS and, in particular, show that our BV master 
equation will really reproduces the Sonoda's WT identity for the 
Wilson action in QED. We will perform this task in Sects. 4 and 5.

\section{Wilson action in the presence of antifields}

We here follow the method and notation developed by Igarashi-Itoh-Sonoda 
(IIS) in Ref.\citen{Igarashi:2007fw}.

Now we consider a general system of Yang-Mills gauge theory and 
denote the action generically as follows by separating the 
kinetic terms from the interaction terms:
\begin{eqnarray}
{\cal S}[\phi] = \frac{1}{2}\phi\cdot D \cdot \phi+ {\cal S}_{I}[\phi].
\label{eq:action}
\end{eqnarray}
Here and henceforth we use the condensed notation for the fields, $\phi^A$,  
with the superfix $A$ standing for all the field indices and 
the matrix notation in momentum space like
\begin{eqnarray}
J \cdot \phi&=& \int_pJ_{A}(-p)\phi^{A}(p), 
\qquad \quad 
\int_p \equiv\int\frac{d^{d}p}{(2\pi)^d}\ ,
\nn\\
\phi\cdot  D  \cdot \phi&=& \int_p 
\phi^{A}(-p) D_{AB}(p) \phi^{B}(p)\ .
\label{eq:cond-not}
\end{eqnarray}
The action (\ref{eq:action}) is understood to contain the gauge fixing 
term and the 
corresponding Faddeev-Popov ghost term. The gauge invariance of the 
system is therefore represented by the invariance under the BRS 
transformation, which we 
\begin{equation}
\delta_\B \phi^A = F^A(\phi) \,.
\end{equation} 
Note that we adopt in this note the convention for this BRS transformation 
$\delta_\B$ to be the operation from the {\em right}: 
$\delta_\B(F\,G)= F\,\delta_\B G +(-)^{\epsilon(G)}(\delta_\B F)\,G$.\footnote{We 
are following this convention by IIS in this note for ease of comparison 
of our results with theirs, although the conversion to the more natural 
convention of left-operation is easy.
} 

We define as usual the generating functional for this system in the 
presence 
of the external sources $J_A$ as well as the BV antifields 
$\phi_A^*$, source functions for the BRS transformation:
\begin{eqnarray}
{\cal Z}_{\phi}[J,\, \phi^*] = \int{\cal D} \phi\,
\exp\left(-{\cal S}[\phi]+ J \cdot \phi- \phi^* \cdot F(\phi)
       \right )\ .
\label{generating-func1}
\end{eqnarray}

To define the Wilson action, we introduce a momentum cutoff 
function $K(p)$ depending only on $p^2$ that behaves as
\begin{eqnarray}
 K(p) \quad  \rightarrow\quad  \left\{
		\begin{array}{ll}
		 1 & (p^2 < \Lambda^2) \\
		  0 & (p^2 \rightarrow\infty)
		\end{array}
               \right. .
\label{cutoff-func}
\end{eqnarray}
We take the function going to $0$ sufficiently rapidly as $p^{2}
\rightarrow\infty$, but $K(p)\not=0$ for any finite $p$ so as for $1/K(p)$ to exist.  
We can now decompose the original fields $\phi^{A}$ into 
the infrared (IR) fields $\Phi^{A}$ and 
the ultraviolet (UV) fields $\tilde\phi^{A}$ 
whose propagators are given by 
$K(p)\left(D_{AB}(p)\right)^{-1}$ and 
$(1-K(p))\left(D_{AB}(p)\right)^{-1}$, respectively: 
\begin{equation}
\phi^A = \Phi^A + \tilde\phi^A\,.
\end{equation}
Remarkably, IIS achieved this task very concisely 
by multiplying the generating functional (\ref{generating-func1}) by 
a gaussian integral over new fields $\theta^{A}$
\begin{equation}
\int{\cal D} \theta\,\exp -\frac{1}{2}
\theta\cdot \frac{D}{K(1-K)} \cdot \theta 
= {\rm constant}
\end{equation} 
and rewriting the $\theta$ fields as 
\begin{eqnarray}
\theta^{A}= (1-K)\Phi^{A}- K\tilde\phi^{A}- (-)^{\epsilon_A}(D^{-1})^{AB}(1-K)J_B\, .
\label{new-fields}
\end{eqnarray} 
Indeed, performing the change of integration variables 
$(\phi^A, \theta^A) \ \rightarrow\ (\Phi^A, \tilde\phi^A)$, we find after a little algebra 
\begin{eqnarray}
{\cal Z}_{\phi}[J,\, \phi^*] &=& N_{J}\int{\cal D} \Phi{\cal D}\tilde\phi\,
\exp-\biggl(\frac{1}{2}\Phi\cdot K^{-1}D \cdot \Phi 
+\frac{1}{2}\tilde\phi\cdot  (1-K)^{-1}D \cdot \tilde\phi 
\nn\\
 &&\hspace{9em}{}+ {\cal S}_{I}[\Phi+ \tilde\phi]
+ \phi^*\cdot F(\Phi+ \tilde\phi)
- J \cdot K^{-1}\Phi 
\biggr),
\end{eqnarray}
where
\begin{eqnarray}
N_{J} \equiv\exp \,{1\over2}(-)^{\epsilon_{A}} J_{A}
 (1-K^{-1})\left(D^{-1}\right)^{AB}J_{B}\,.
\end{eqnarray} 
Note that the external sources $J_A$ no longer couple to the UV 
fields $\tilde\phi^A$ and do to the IR fields $\Phi^A$ with the factor 
$K^{-1}$. In this sense $K^{-1}J_A$ play the role of the low-energy 
source functions 
for the IR fields $\Phi^A$. We will shortly see that 
it is suitable to define the antifields $\Phi^*_A$ in the low-energy 
world as
\begin{equation}
\Phi^*_A = K^{-1}\phi^*_A\,.
\label{IRantifield}
\end{equation}

The Wilson action for the IR fields $\Phi^A$ is thus defined in the 
presence of the BV antifields $\Phi^*_A = K^{-1}\phi^*$ by
\begin{equation}
S[\Phi,\,\Phi^*] \equiv{1\over2}\Phi\cdot K^{-1}D \cdot \Phi+ S_{I}[\Phi,\,\Phi^*]
\label{Wilsonaction1}
\end{equation}
with $S_{I}[\Phi,\,\Phi^*]$ given by 
\begin{eqnarray}
e^{-S_{I}[\Phi,\,\Phi^*]} \equiv\int{\cal D}\tilde\phi\,
\exp - \Bigl(\frac{1}{2} \tilde\phi\cdot (1-K)^{-1}D \cdot \tilde\phi+ 
{\cal S}_{I}[\Phi+\tilde\phi] + K\Phi^* \cdot\,F(\Phi+\tilde\phi)
\Bigr)\,.
\label{Wilsonaction2}
\end{eqnarray} 
The generating functional in the low-energy world is given in terms of 
this Wilson action by 
\begin{eqnarray}
Z_{\Phi}[K^{-1}J,\, \Phi^*] = \int{\cal D} \Phi\,
\exp\left(-S[\Phi,\,\Phi^*]+K^{-1}J \cdot \Phi\right)\,,
\label{generating-func2}
\end{eqnarray}
which is related to the original one for $\phi$ by
\begin{eqnarray}
{\cal Z}_{\phi}[J,\, \phi^*] = N_{J} Z_{\Phi}[K^{-1}J,\, \Phi^*]\,.
\label{gf-relation}
\end{eqnarray}

\section{BV master equation for the Wilson action}

It is now easy to prove that the Wilson action 
$S[\Phi,\,\Phi^*]$ thus defined in (\ref{Wilsonaction1}) with 
(\ref{Wilsonaction2}) 
satisfies the quantum BV master equation. 

In the generating functional ${\cal Z}_{\phi}$ in (\ref{generating-func1}),
we perform the change of integration variables 
\begin{equation}
\phi^A \ \rightarrow\ \phi^A + F^A(\phi)\lambda 
\end{equation}
where the field shift $\delta\phi^A$ is the same as the BRS transformation 
$\delta_\B\phi^A$ multiplied by a Grassmann-odd parameter $\lambda$. 
Since the original action 
${\cal S}[\phi]$ as well as BV antifield terms $\phi^*\cdot F(\phi)$ 
are BRS invariant, $\delta_\B{\cal S}[\phi]=0$ and 
$\delta_\B(\phi^*\cdot  F(\phi)) = \phi^*\cdot \delta_\B(\delta_\B\phi)=0$, 
the only changes come from the external source terms $J\cdot \phi$ 
and so we obtain 
\begin{equation}
\int{\cal D} \phi\, \bigl(J\cdot F(\phi)\bigr)
\exp\left(-{\cal S}[\phi]+ J \cdot \phi- \phi^* \cdot F(\phi)
       \right )=0 \quad  \rightarrow \quad 
J\cdot {\delta^l\over\delta\phi^*}{\cal Z}_{\phi}[J,\, \phi^*] = 0\,,
\end{equation}
where the notations $\delta^l$ and $\delta^r$ distinguish the operations from 
the left and right, respectively, when necessary.
Using the relations (\ref{IRantifield}), (\ref{generating-func2}) and (\ref{gf-relation}), we can rewrite this into
\begin{equation}
J\cdot K^{-1}{\delta^l\over\delta\Phi^*}Z_{\Phi}[K^{-1}J,\, \Phi^*] 
=
\VEV{J\cdot K^{-1}{\delta^l S\over\delta\Phi^*}}_{K^{-1}J,\ \Phi^*}
= 0\,,\label{WTidentity}
\end{equation}
where use has been made of the notation:
\begin{equation}
\VEV{ \ \cdots\  }_{K^{-1}J,\ \Phi^*}
\equiv 
\int{\cal D} \Phi\,
\left( \ \cdots\ \right)
e^{\left(-S[\Phi,\,\Phi^*]+K^{-1}J \cdot \Phi\right)}\ .
\end{equation}

On the other hand, since integration of any  total derivative 
quantity is vanishing  we have 
\begin{eqnarray}
0 &=& \int{\cal D} \Phi\,
{\delta^r\over\delta\Phi^A}\left({\delta^l S\over\delta\Phi^*_A}
e^{\left(-S[\Phi,\,\Phi^*]+K^{-1}J \cdot \Phi\right)} \right) \nn\\
&=& \VEV{{\delta^r\delta^lS\over\delta\Phi^A\delta\Phi^*_A} - {\delta^r S\over\delta\Phi^A}{\delta^l S\over\delta\Phi^*_A} 
+ K^{-1}J \cdot{\delta^l S\over\delta\Phi^*} }_{K^{-1}J,\ \Phi^*}. 
\end{eqnarray}
The above identity (\ref{WTidentity}), therefore, yields
\begin{equation}
\VEV{{\delta^r\delta^lS\over\delta\Phi^A\delta\Phi^*_A} - {\delta^r S\over\delta\Phi^A}{\delta^l S\over\delta\Phi^*_A} 
 }_{K^{-1}J,\ \Phi^*} =0\,.
\end{equation}
This identity holds for any values of the external source functions 
$J_A$, and so implies that the inside quantity itself vanishes:
\begin{equation}
{\delta^r S\over\delta\Phi^A}{\delta^l S\over\delta\Phi^*_A} =
{\delta^r\delta^lS\over\delta\Phi^A\delta\Phi^*_A}\,.
\label{BVmastereq}
\end{equation} 
This is the quantum BV master equation for the Wilson action 
$S[\Phi,\,\Phi^*]$ (\ref{MasterAction}) in the presence of antifield background.
It is quite remarkable that exactly the same form of BV master equation as
that for the continuum theory without UV cutoff holds here for 
the Wilson effective action for 
the IR fields. In this sense, the UV momentum cutoff (smooth momentum 
cutoff, at least) seems compatible with the general non-Abelian 
gauge-invariance, as claimed by Sonoda in QED case.\cite{Sonoda:2007dj}

\section{The Sonoda's WT identity for Wilson action in QED case}

Up to here we have been considering 
the general non-Abelian gauge theories. 
For the rest of this note, however, we consider the Abelian gauge theories. 
The special circumstance in the Abelian case is that the Faddeev-Popov 
ghost and antighost are free from interaction in the usual covariant gauges. 
This fact makes the BV antifield dependence of the 
master action considerably explicit and enables us to make connections with 
the previous works in QED case by Sonoda\cite{Sonoda:2007dj} and by 
IIS.\cite{Igarashi:2007fw}

We now show that our quantum BV Master equation 
reproduces the Sonoda's WT identity for the Wilson action in the 
case of QED. We also show that our Wilson action in the presence of the 
antifields indeed deduces the IIS's BV master action in 
Ref.\citen{Igarashi:2007fw} which they elaborated starting 
from the Sonoda's WT identity. The master action constructed by them is 
not necessarily equal to our master action, although the Wilson action 
in the absence of the antifields is of course unique. As was already 
noted by IIS, there is an ambiguity of performing canonical 
transformations in the field and antifield variable space. In the next 
section, we will explicitly give the canonical transformation which 
transforms our variables to theirs, and show that our master action 
indeed deduces their master action exactly.

The fields, antifields and external sources in the QED case are given by
\begin{eqnarray}
\phi^A &=& \{ a_\mu,\ b,\ c,\ \bar c, \psi,\ \bar\psi\}\,, \nn\\
\phi^*_A &=& \{ a^*_\mu,\ b^*,\ c^*,\ \bar c^*, \psi^*,\ \bar\psi^*\}\,, \nn\\
J_A &=& \{ J_\mu,\ J_b,\ J_c,\ J_{\bar c}, J_\psi,\ J_{\bar\psi}\}\,.
\end{eqnarray}
But the antifields $b^*,\ c^*$ actually do not appear in QED case since 
$\delta_\B b=\delta_\B c=0$.\footnote{Since the antifield 
variables $b^*,\ c^*$ are missing, the fields 
$b$ and $c$ are no longer the canonical field variables but becomes 
mere `parameters' in the BV field-antifield formalism in the next section.} 
We denote the IR fields $\Phi^A$ by the corresponding upper case letters: 
\begin{equation}
\Phi^A = \{ A_\mu,\ B,\ C,\ \bar C, \Psi,\ \bar\Psi\}\, . 
\end{equation}
The action (\ref{eq:action}) in the covariant $\xi$ gauge reads explicitly 
\begin{eqnarray}
\frac{1}{2}\phi\cdot D \cdot \phi 
&=& 
\int_k \left[{1\over2}a_\mu(-k)\bigl(k^2\delta_{\mu\nu}-k_\mu k_\nu\bigr)a_\nu(k) 
+ \bar c(-k)ik^2c(k) \right. \nn\\
&& \left.{}-b(-k)\bigl(ik^\mu a_\mu(k)+{\xi\over2}b(k)\bigr) \right]
+ \int_p \bar\psi(-p)\bigl(\pslash +m\bigr)\psi(p) \, , \nn\\
{\cal S}_{I}[\phi] &=& \int_{p,k} -e\,\bar\psi(-p-k)\gamma^\mu\psi(p)a_\mu(k)\,.
\end{eqnarray}
The antifield terms $\phi^*_AF^A(\phi)$ are given by
\begin{eqnarray}
\phi^*_AF^A(\phi) &=& 
\int_k\left[a_\mu^*(-k)(-ik^\mu c(k)) + \bar c^*(-k)ib(k)\right] \nn\\
&&\ {}-ie\int_{p,k}\left[ \psi^*(-p)\psi(p-k)c(k) + 
c(k)\bar\psi(p-k)\bar\psi^*(-p)\right]\,.
\end{eqnarray}

To make explicit the antifield dependence of the Wilson action, 
we proceed as follows. First, the antifield dependence solely comes  
from the antifield term $K\Phi^*\cdot F(\Phi+\tilde\phi)$ which is contained 
in the exponent of the integrand of the $\tilde\phi^A$ integration 
defining $S_I[\Phi,\,\Phi^*]$ in (\ref{Wilsonaction2}).
That term is expanded in powers of the integration variables $\tilde\phi^A$:
\begin{equation}
K\Phi^*\cdot F(\Phi+\tilde\phi)=
K\Phi^*\cdot \left[F(\Phi) + F'(\Phi)\tilde \phi 
+ {1\over2}F''\tilde\phi\tilde\phi\right]\,.
\label{antifieldterm}
\end{equation}
Among the integration variables $\tilde\phi^A$, the FP ghost variable 
$\tilde c$ can be set equal to zero in QED case. 
This is because FP ghosts 
are free and the antighost variable $\tilde{\bar c}$ appears only in the 
free kinetic term $\tilde{\bar c}(-k) (1-K)^{-1}ik^2 \tilde c(k)$ so that 
any terms proportional to $\tilde c$ can be absorbed by shifting the 
antighost variable $\tilde{\bar c}$. 
By this procedure, all the quadratic terms and a part of the linear terms 
in $\tilde \phi$ are eliminated here in Eq.~(\ref{antifieldterm}).\footnote{
In the usual covariant gauges, all the quadratic terms in $\tilde{\phi}$ contain the FP
ghost $\tilde{c}$ as one of the two $\tilde{\phi}$'s. The fact that all the
quadratic terms
can be eliminated in QED gives the reason why all the antifield
dependences there
can be represented by the shift of the fields in the Wilson action.
The linear terms
in $\tilde{\phi}$ can be eliminated by the shift of the UV fields which turns to give
the shift of the IR fields in the Wilson action, as we will show shortly.
In non-Abelian cases, however, the FP ghosts are interacting fields and so the
quadratic terms in $\tilde{\phi}$ cannot be eliminated, implying that the antifield
dependences exit which cannot be represented by the shift of the fields.
} 
The remaining linear terms in $\tilde\phi^A$ are now for 
$\tilde\phi^A=\{ \tilde b,\ \tilde\psi,\ \tilde{\bar\psi} \}$. They can be 
absorbed into the kinetic terms $(1/2)\tilde\phi\cdot(1-K)^{-1}D\cdot\tilde\phi$ by 
shifting the integration variables 
\begin{equation}
\tilde\phi^{\prime A} = \tilde\phi^A + f^{AB}(\Phi)\Phi^*_B
\end{equation}
with the shift $(f(\Phi)\cdot\Phi^*)^A$ determined by the condition
\begin{equation}
K\Phi^*\cdot F'(\Phi)\tilde\phi= (f(\Phi)\cdot\Phi^*)\cdot (1-K)^{-1}D\cdot\tilde\phi\ .
\label{eq:condition}
\end{equation}
Then the expression for the $S_{I}[\Phi,\,\Phi^*]$ in (\ref{Wilsonaction2}) 
now becomes 
\begin{eqnarray}
\exp(-S_{I}[\Phi,\,\Phi^*]) &=& 
\exp - \Bigl(S'_{I}[\Phi']+K\Phi^*\cdot F(\Phi) 
-\frac{1}{2} (f(\Phi)\cdot\Phi^*)\cdot (1-K)^{-1}D 
\cdot(f(\Phi)\cdot\Phi^*)
\Bigr) \nn\\
\exp(-S'_{I}[\Phi']) &=& 
\int{\cal D}\tilde\phi'\,
\exp - \Bigl(\frac{1}{2} \tilde\phi'\cdot (1-K)^{-1}D \cdot \tilde\phi'+ 
{\cal S}_{I}[\Phi'+\tilde\phi'] \Bigr)\ .
\label{eq:Sprime}
\end{eqnarray}
Here in the argument of the interaction term 
${\cal S}_{I}[\Phi'+\tilde\phi']$, we have also shifted the IR fields 
\begin{equation}
\Phi^{\prime A}= \Phi^A - (f(\Phi)\cdot\Phi^*)^A \,,
\end{equation}
such that the 
arguments remain intact, $\Phi+\tilde\phi= \Phi'+\tilde\phi'$.
At this stage $S'_{I}[\Phi']$ depends on the antifields only through 
the shifted IR variables $\Phi^{\prime A}$. 
Explicitly these field shifts are given by
\begin{eqnarray}
A_\mu'(k) &=& A_\mu(k) + {k_\mu\over k^2}\bigl(1-K(k)\bigr)K(k)\bar C^*(k), \nn\\
\Psi'(p) &=& \Psi(p) -ie {1-K(p)\over\pslash+m}\int_k K(p-k)\bar\Psi^*(p-k)C(k), \nn\\
\bar\Psi'(-p) &=& \bar\Psi(-p) 
-ie \int_k K(p+k)\Psi^*(-p-k)C(k){1-K(p)\over\pslash+m}\ .
\label{eq:primedfield}
\end{eqnarray}
The other fields $B,\ C$ and $\bar C$ remain intact. 

Now, we should also do the rewriting $\Phi\rightarrow\Phi'+(f(\Phi)\cdot\Phi^*)$ in all
the other terms in the Wilson action
\begin{equation}
S[\Phi,\,\Phi^*] = {1\over2}\Phi\cdot K^{-1}D \cdot \Phi 
+K\Phi^*\cdot F(\Phi) 
-\frac{1}{2} (f(\Phi)\cdot\Phi^*)\cdot (1-K)^{-1}D 
\cdot(f(\Phi)\cdot\Phi^*)
+S'_{I}[\Phi']\ .
\end{equation}
We need to rewrite the kinetic terms $(1/2)\Phi\cdot K^{-1}D \cdot \Phi$ 
and the antifield terms $K\Phi^*\cdot F(\Phi)$, but $f(\Phi)$ need not be 
rewritten since it contains only $C$. 
We have
\begin{eqnarray}
S[\Phi,\,\Phi^*] &=& {1\over2}\Phi'\cdot K^{-1}D \cdot \Phi'+S'_{I}[\Phi'] \nn\\ 
&&{}+\hbox{(linear terms in $\Phi^*$)} +
\hbox{(quadratic terms in $\Phi^*$)}\ ,
\label{MasterAction}
\end{eqnarray}
where the (linear terms in $\Phi^*$) is given after a short computation by
\begin{eqnarray}
&&\hspace{-3em}\hbox{(linear terms in $\Phi^*$)} \nn\\
&=&
K\Phi^*\cdot F(\Phi') + \Phi'\cdot K^{-1}D \cdot (f(\Phi)\cdot\Phi^*) 
  \nn\\
&=& 
\int_k \Bigl( K(k) A^*_\mu(-k)\bigl(-ik^\mu C(k)\bigr) 
+ \bar C^*(-k)iB(k) \Bigr) \nn\\
&&{}+ie\int_{p,k}\left(
K(p)\Psi^*(-p)C(k){\Psi'(p-k)\over K(p-k)} 
+ {\bar\Psi'(p-k)\over K(p-k)} K(p)\bar\Psi^*(-p)C(k)
\right) 
\label{eq:linearterm}
\end{eqnarray}
and the (quadratic terms in $\Phi^*$) is
\begin{eqnarray}
&&\hspace{-3em}\hbox{(quadratic terms in $\Phi^*$)} \nn\\
&=&
K\Phi^*\cdot F'(\Phi')(f(\Phi)\cdot\Phi^*) 
+\frac{1}{2} (f(\Phi)\cdot\Phi^*)\cdot \left[-{1\over1-K}+{1\over K}\right] D 
\cdot(f(\Phi)\cdot\Phi^*) \ .
\end{eqnarray}
Here we can set $F'(\Phi')=F'(\Phi)$ 
since only fermion terms are relevant and $\Phi=C$ appears there. 
Therefore, 
using the relation (\ref{eq:condition}),
we find 
\begin{equation}
K\Phi^*\cdot F'(\Phi)(f(\Phi)\cdot\Phi^*) 
= (f(\Phi)\cdot\Phi^*)\cdot (1-K)^{-1}D\cdot
(f(\Phi)\cdot\Phi^*)\,,
\end{equation}
so that the quadratic terms become
\begin{equation}
\hbox{(quadratic terms in $\Phi^*$)}=
\frac{1}{2} (f(\Phi)\cdot\Phi^*)\cdot \left[{1\over1-K}+{1\over K}\right] D 
\cdot(f(\Phi)\cdot\Phi^*)\,.
\end{equation}
The shift $(f(\Phi)\cdot\Phi^*)$ for the gauge field $A_\mu$ yields no 
quadratic term in $\Phi^*$ here since it is proportional to $ik_\mu$ while 
the kinetic operator $D$ is transversal $\propto(\delta_{\mu\nu}k^2-k_\mu k_\nu)$.
Thus only the fermion kinetic term is relevant and we find
the explicit form for the (quadratic terms in $\Phi^*$):
\begin{eqnarray}
&=&(f(\Phi)\cdot\Phi^*)^{\bar\Psi}(-p){\pslash+m\over K(p)\bigl(1-K(p)\bigr)}(f(\Phi)\cdot\Phi^*)^\Psi(p) \nn\\
&=&
(ie)^2\int_{p,k,l}
K(p+k)\Psi^*(-p-k)C(k){1-K(p)\over K(p)(\pslash+m)}K(p-l)\bar\Psi^*(p-l)C(l)\,.
\label{eq:quadraticterm}
\end{eqnarray} 

Now that we have made explicit the antifield dependence of 
the Wilson action, we can easily see that our BV master equation 
(\ref{BVmastereq}) really reproduces the Sonoda's Ward identity for 
the Wilson action by putting all the antifield equal to zero after 
evaluating the antifield derivatives. 

\section{Quantum BV master equation by IIS}

Our master action $S[\Phi,\,\Phi^*]$, however, still does not coincide 
with that derived by IIS, 
although both reproduce the same Sonoda equation when the antifields 
are set equal to zero. This is due to the fact that there is a freedom 
of doing canonical transformation in the field and antifield space. 

The clear differences of the two actions are 
that the implicit antifield dependence in our case exists through
three primed variables $A_\mu',\ \Psi',\ \bar\Psi'$  in (\ref{eq:primedfield}), 
while it is only through $\bar\Psi'$ in the IIS case, and also the 
shifts in $\bar\Psi'$ do not coincides with each other. 
We are thus lead to performing the canonical transformation which 
identify our first two primed variables $A_\mu',\ \Psi'$ with IIS's 
variables $A_{\mu\,\IIS},\ \Psi_\IIS$. 
The generating function(al) of the desired canonical transformation
is easily found to be 
\begin{eqnarray}
&&\hspace{-2em}W[\Phi,\,\Phi^*_{\IIS}] \nn\\
&=&\int_k\left[
A^{*\mu}_\IIS(-k)\Bigl( A_\mu(k)+{k_\mu\over k^2}K(k)(1-K(k))\bar C^*_\IIS(k)
 \Bigr)
+\bar C^*_\IIS(-k) \bar C(k)\right] \nn\\
&&{}+\int_p\left[\Psi^*_\IIS(-p) \Bigl(\Psi(p) 
-ie(1-K(p))\int_k (\pslash+m)^{-1}K(p-k)\bar\Psi^*_\IIS(p-k)C(k)\Bigr)\right.
 \nn\\
&&{}\hspace{4em}{}+\bar\Psi(p)\bar\Psi^*_\IIS(-p) \biggr]\, .
\end{eqnarray}
Then, actually, the canonical variable relation
$\Phi^*_A(k) =\delta^lW[\Phi,\,\Phi^*_{\IIS}]/\delta\Phi^A(-k)$ says first 
that all the antifield variables coincide
\begin{eqnarray}
A_\mu^*(k) &=& A^*_{\mu\,\IIS}(k)\,, \qquad 
\bar C^*(k) = \bar C^*_\IIS(k)\,, \nn\\
\Psi^*(p) &=& \Psi^*_\IIS(p)\,, \qquad  \ \ \,  
\bar\Psi^*(p) = \bar\Psi^*_\IIS(p)\,,
\end{eqnarray}
and 
$\Phi^A_\IIS(k) =\delta^lW[\Phi,\,\Phi^*_{\IIS}]/\delta\Phi^*_{A\,\IIS}(-k)$ 
gives the relation of the field variables: 
\begin{eqnarray}
A_{\mu\,\IIS}(k) &=& A_\mu(k)+{k_\mu\over k^2}K(k)(1-K(k))\bar C^*_\IIS(k) 
= A'_\mu(k)\,,
\label{eq:A}\\
\bar C_\IIS(k) &=& \bar C(k) - K(k)(1-K(k)){k_\mu\over k^2}A^{*\mu}_\IIS(k)\,, 
\label{eq:barC}\\
\Psi_\IIS(p) &=& \Psi(p) 
-ie(1-K(p))\int_k (\pslash+m)^{-1}K(p-k)\bar\Psi^*_\IIS(p-k)C(k) 
=\Psi'(p)\, ,
\label{eq:Psi}
\\
\bar\Psi_\IIS(-p) &=& \bar\Psi(-p) 
-ie(1-K(p+k))
\int_k \Psi^*_\IIS(-p-k) (\pslash+\kslash+m)^{-1}K(p)C(k)
\nn \\
&=& \bar\Psi'(-p) + ie \int_k \Psi^*_\IIS(-p-k) C(k) U(-p-k,\,p)\,, 
\label{eq:barPsi}\\
{\rm with} &&U(-p-k,\,p)= {1-K(p)\over\pslash+m}K(p+k)
- {1-K(p+k)\over\pslash+\kslash+m}K(p)\,. 
\label{eq:U}
\end{eqnarray}
The first and third equations, 
(\ref{eq:A}) and (\ref{eq:Psi}), give the desired transformation for the 
photon and electron variables. The second equation (\ref{eq:barC}) gives  
the `unexpected' transformation for antighost $\bar C$ 
associated with the canonical transformation of the photon field. 
This shift of $\bar C$ 
in the FP ghost kinetic term $K^{-1}(k)\bar C(-k) ik^2 C(k)$ in the Wilson action, 
yields an additional term 
\begin{equation}
A_\mu^*(-k) \bigl(1-K(k)\bigr)\bigl(-ik^\mu C(k)\bigr)\,,
\end{equation}
which transforms the non-canonical weight $K(k)$ in the already existing 
antifield term $K(k) A^*_\mu(-k)\bigl(-ik^\mu C(k)\bigr)$ in 
(\ref{eq:linearterm}) into the canonical weight 1. 

Eq.~(\ref{eq:barPsi}) indicates that our primed field $\bar\Psi'$ 
agrees with IIS's primed field $\bar\Psi'$ 
since the function $U(-p-k,\,p)$ here coincides 
with theirs. So we have to rewrite the $\bar\Psi'$ variable in the 
$\bar\Psi'\bar\Psi^*C$ term in (\ref{eq:linearterm}) 
in terms of $\bar\Psi_\IIS$. The shift proportional to $U$ there 
yields an additional contribution to the (quadratic terms in $\Phi^*$):
\begin{equation}
e^2\int_{p,k,l}
{K(-p+l)\over K(p)} \Psi^*(-p-k)C(k)U(-p-k,\,p)\bar\Psi^*(p-l)C(l)
\,.
\end{equation}  
As shown in Eq.~(\ref{eq:U}), the function $U(-p-k,\,p)$ contains two 
terms. The contribution from the first term $\propto(\pslash+m)^{-1}$ 
to this exactly cancels the (quadratic terms in $\Phi^*$) in 
(\ref{eq:quadraticterm}), thus leaving the contribution from the
second term
\begin{equation}
-e^2\int_{p,k,l}
\Psi^*(-p-k)C(k)
{1-K(p+k)\over\pslash+\kslash+m}K(p-l)\bar\Psi^*(p-l)C(l)\,.
\end{equation}
This term, however, vanishes by itself. Indeed, 
by shifting the momentum $p\rightarrow p-k$, it is rewritten into the form 
\begin{equation}
-e^2\int_{p,k,l}
\Psi^*(-p)
{1-K(p)\over\pslash+m}K(p-k-l)\bar\Psi^*(p-k-l)C(k)C(l) \,,
\end{equation}
which clearly vanishes since $C(k)C(l)$ is antisymmetric under the 
exchange of $k$ and $l$ while they appear symmetrically in the form 
$p-k-l$ in the arguments of the other functions.
Thus the quadratic terms in $\Phi^*$ vanish completely.

Now collecting the terms altogether and writing the Wilson action in 
terms of the IIS field variables $\Phi_\IIS$ but omitting the index 
$\IIS$ for simplicity, 
we find
\begin{eqnarray}
S[\Phi,\,\Phi^*] &=& {1\over2}\Phi'\cdot K^{-1}D \cdot \Phi'+S'_{I}[\Phi'] \nn\\ 
&&{}+\int_k \Bigl( A^*_\mu(-k)\bigl(-ik^\mu C(k)\bigr) 
+ \bar C^*(-k)iB(k) \Bigr) \\
&&{}+ie\int_{p,k}\left(
K(p)\Psi^*(-p)C(k){\Psi(p-k)\over K(p-k)} 
+ {\bar\Psi(p-k)\over K(p-k)} K(p)\bar\Psi^*(-p)C(k)
\right)\, \nn
\end{eqnarray}
with $S'_{I}[\Phi']$ defined in Eq.~(\ref{eq:Sprime}). 
Here the primed field $\Phi^{\prime\, A}$ denotes that only the 
$\bar\Psi$ is shifted:
\begin{eqnarray}
\Phi^{\prime\, A} &=& \{ A_\mu,\ B,\ C,\ \bar C,\ \Psi,\ \bar\Psi'\}\,, \nn\\
\bar\Psi'(-p)
&=& \bar\Psi(-p) - ie \int_k \Psi^*_\IIS(-p-k) C(k) U(-p-k,\,p) \,.
\end{eqnarray} 
with $U(-p-k,\,p)$ given in Eq.~(\ref{eq:U}). 
This expression for the master action exactly agrees with that derived 
by IIS. [See Eqs.(41) and (42) in Ref.~\citen{Igarashi:2007fw}.]

\section{Summary and Discussion}

In this paper, we have derived the BV master equation for the Wilson 
action in the general non-Abelian gauge theory following IIS's 
work\cite{Igarashi:2007fw}. We have introduced the antifields as the 
sources for the BRS transformation of the fields from the starting action. 
In QED case, we have 
made explicit the antifield dependence of our master action, and have 
shown that our master action deduces the IIS 's master action and 
the two expressions exactly agree with each other via a canonical 
transformation in the field and antifield variable space, 
whose generating function
was found explicitly. 
In particular, our BV master equation reproduced the Sonoda's WT identity 
when the antifields are set equal to zero.

In this approach, the BRS transformation for the IR field $\Phi$ is defined 
as follows by the master action $S[\Phi,\Phi^*]$ itself:
\begin{eqnarray}
\delta_Q \Phi=(\Phi, S[\Phi,\Phi^*])-\Delta\Phi,
\end{eqnarray}
where $(X,Y)$ and $\Delta$ are:
\begin{eqnarray}
(X,Y)&\equiv&
\frac{\partial^r X}{\partial\Phi^A}\frac{\partial^l Y}{\partial\Phi^*_A}
-\frac{\partial^r X}{\partial\Phi^*_A}\frac{\partial^l Y}{\partial\Phi^A}\,,\\
\Delta&\equiv&
(-)^{\epsilon_A +1}\frac{\partial^r}{\partial\Phi^A}\frac{\partial^r}{\partial\Phi^*_A}\,.
\end{eqnarray}
The antifield dependence of the master action is non-trivial 
in the general non-Abelian gauge theories, and so is this BRS 
transformation. 
The explicit form of the BRS transformation is thus determined 
simultaneously as that of the Wilson action. 

The Polchinski equation (exact renormalization group equation)
for the master action is invariant under this 
``quantum" BRS transformation, and we can obtain the BRS invariant 
renormalization group flows for the Wilson effective action of the 
gauge theories.

\begin{center}
{\bf Note added}
\end{center}
After submitting this note to the arXive, we became aware of the work
by Igarashi,
Itoh and So, Ref.\citen{Igarashi:1999rm}, in which the authors already derived the exact quantum
BV master equation for the Wilson action in the presence of
antifields. Although
their IR fields are defined to be ``average fields" differently from
ours, the BV master
equation holds very similar to ours.

\section*{Acknowledgements}
We would like to thank Yuji Igarashi, Katsumi Itoh and Hidenori Sonoda 
for helpful discussions. One of the authors (T.K.) is partially supported 
by a Grant-in-Aid for Scientific Research (B) No.\ 16340071 from Japan 
Society for the Promotion of Science. The authors E.I.\ and T.K.\ are 
also supported by a Grant-in-Aid for the 21st Century COE ``Center for 
Diversity and Universality in Physics".

%

\end{document}